\documentclass[aps,prl,twocolumn,superscriptaddress]{revtex4-1}

\usepackage{gensymb}
\usepackage{bm}
\usepackage{color}
\usepackage{graphicx}
\usepackage[colorlinks=true,linkcolor=blue]{hyperref}%
\usepackage{amsmath}
\expandafter\ifx\csname package@font\endcsname\relax\else
 \expandafter\expandafter
 \expandafter\usepackage
 \expandafter\expandafter
 \expandafter{\csname package@font\endcsname}%
\fi
\begin{document}

\title{Generation of high-energy electron-positron beams in the collision of a laser-accelerated electron beam and a multi-petawatt laser}

\author{M. Lobet}
\email{mathieu.lobet@cea.fr}
\affiliation{CEA, DAM, DIF, F-91297, Arpajon, France}
\affiliation{CELIA, UMR 5107, Universit\'e de Bordeaux-CNRS-CEA, 33405, Talence}
\author{X. Davoine}
\email{xavier.davoine@cea.fr}
\affiliation{CEA, DAM, DIF, F-91297, Arpajon, France}
\author{E. d'Humi\`eres}
\affiliation{CELIA, UMR 5107, Universit\'e de Bordeaux-CNRS-CEA, 33405, Talence}
\author{L. Gremillet}
\email{laurent.gremillet@cea.fr}
\affiliation{CEA, DAM, DIF, F-91297, Arpajon, France}

\begin{abstract}
Generation of antimatter via the multiphoton Breit-Wheeler process in an all-optical scheme will be made possible on forthcoming high-power laser facilities
through the collision of wakefield-accelerated GeV electrons with a counter-propagating laser pulse with $10^{22}-10^{23}\,\mathrm{Wcm}^{-2}$ peak intensity.
By means of integrated 3D particle-in-cell simulations, we show that the production of positron beams with $0.05-1\,\mathrm{nC}$ total charge,
$100-400\,\mathrm{MeV}$ mean energy and $0.01-0.1\,\mathrm{rad}$ divergence is within the reach of soon-to-be-available laser systems. The variations of
the positron beam's properties with respect to the laser parameters are also examined. 
\end{abstract}

\maketitle



\textit{Introduction} - The generation of dense positron beams is of great interest in many research areas, encompassing fundamental science \cite{RUFFINI_PR_487_1_2010},
accelerator physics \cite{CORDE_NATURE_524_442_2015}, material analysis \cite{TUOMISTO_RMP_85_1583_2013} and laboratory astrophysics \cite{MESZAROS_AA_40_137_2002}.
In the latter field, a driving goal is to understand the formation and dynamics of the electron-positron ($e^-e^+$) pair plasmas involved in a number of powerful
space environments (pulsar winds, gamma-ray bursts, active galactic nuclei) \cite{BYKOV_AAR_19_42_2011}. The laboratory reproduction of these phenomena is hampered by the
need to create pair plasmas dense enough to trigger collective effects. The use of ultra-intense, short-pulse lasers provides a promising path to this objective,
as testified by the various schemes put forward in recent years
\cite{CHEN_PRL_102_105001_2009,*CHEN_PRL_114_215001_2015,SARRI_Nature_2015,SHEN_PRE_65_016405_2001,BELL_PRL_101_200403_2008,NERUSH_PRL_106_035001_2011,PIKE_NATURE_8_434_2014}.
The experimental studies carried out so far exclusively rely upon the Bethe-Heitler conversion of bremsstrahlung \mbox{$\gamma$-rays} emitted by energetic electrons
injected through thick ($\sim \mathrm{cm}$) high-$Z$ targets. These electrons are produced either by direct irradiation of the convertor target by an intense
($\sim 10^{20}\,\mathrm{Wcm}^{-2}$) picosecond laser \cite{CHEN_PRL_102_105001_2009,*CHEN_PRL_114_215001_2015} or by a few-femtosecond laser wakefield accelerator
\cite{SARRI_Nature_2015}, in both cases leading to record positron densities of $\sim 10^{16}\,\mathrm{cm}^{-3}$.    

The coming into operation of multi-PW lasers (e.g., the CILEX-Apollon \cite{APOLLON} and ELI \cite{ELI} facilities) expected to reach intensities $\sim 10^{23}\,\mathrm{Wcm}^{-2}$, should make it possible to investigate alternative concepts of pair generation, based on the Breit-Wheeler decay of $\gamma$-rays produced by nonlinear Compton
scattering \cite{BELL_PRL_101_200403_2008,DIPIAZZA_RMP_84_1177_2012}. Theory and simulations predict that  quasi-neutral, high-density ($>10^{21}\,\mathrm{cm}^{-3}$)
relativistic pair plasmas could be generated from gas jets or thin solid foils \cite{RIDGERS_PRL_108_165006_2012,BRADY_PRL_109_245006_2012}, through pair cascading
from seed electrons or photons \cite{NAROZHNY_JETP_80_382_2004,BULANOV_PRL_105_220407_2010,NERUSH_PRL_106_035001_2011,WU_PRD_90_013009_2014}, or using the
flying-mirror concept  \cite{BULANOV_NIMPRSA_660_31_2011}. In the near future, however, the most accessible route to laser-driven Breit-Wheeler pair production will
exploit the collision of relativistic electrons with counter-propagating high-power lasers \cite{TUCHIN2010,SOKOLOV_PRL_105_195005_2010,BULANOV_PRA_87_06210_2013}.
This concept was demonstrated two decades ago at SLAC by making a 46 GeV electron beam interact with a $10^{18}\,\mathrm{Wcm}^{-2}$ laser \cite{BURKE1997},
and is planned to be reproduced at higher electron energies on the future linear collider \cite{HARTIN2014}. An all-optical scheme, based on a laser wakefield
accelerator (LWFA) instead of a conventional accelerator, should be within the reach of multi-PW, multi-beam laser systems. As a first step, consistently with theoretical
predictions \cite{MACKENROTH_PRA_032106_2011,NEITZ_PRL_111_054802_2013,VRANIC_PRL_113_134801_2014,BLACKBURN_PRL_112_015001_2014,*BLACKBURN_PPCF_57_075002_2015,
HARVEY_SPIE_950908_2015}, recent state-of-the-art experiments have confirmed the potential of this configuration for generating high-brilliance $\gamma$-ray
photon beams \cite{SARRI_PRL_113_224801_2014}.  In order to provide guidelines for future experiments, we present in this Letter the first integrated,
one-to-one simulation study of the pair production  expected on the upcoming multi-PW laser facilities. Focusing on interaction conditions accessible to
the CILEX-Apollon system \cite{APOLLON}, we characterize in detail the generated $e^-e^+$ beam and examine the sensitivity of its properties to the laser parameters.  

\begin{figure}[t]
\begin{center}
\includegraphics[width=0.235\textwidth]{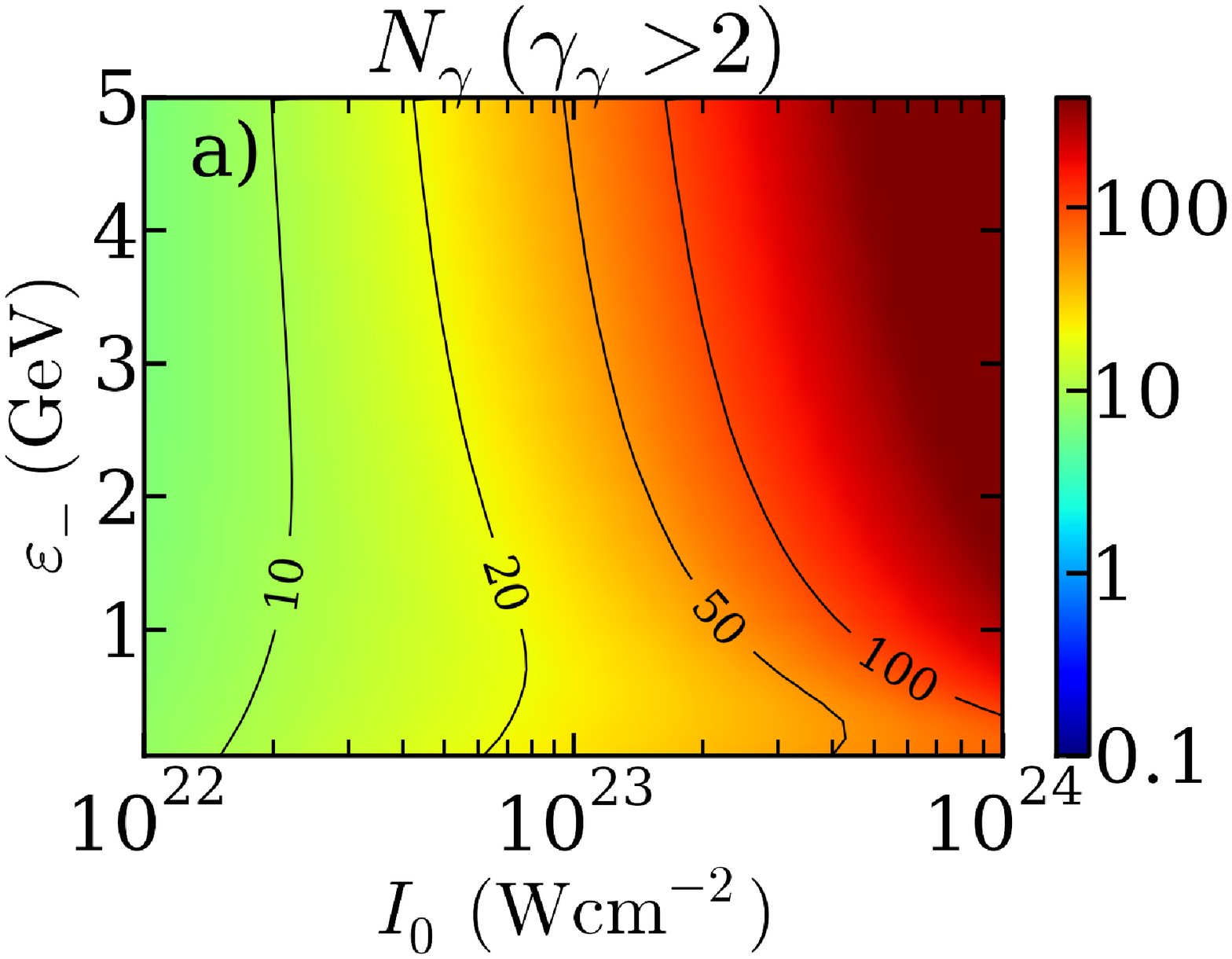}
\includegraphics[width=0.235\textwidth]{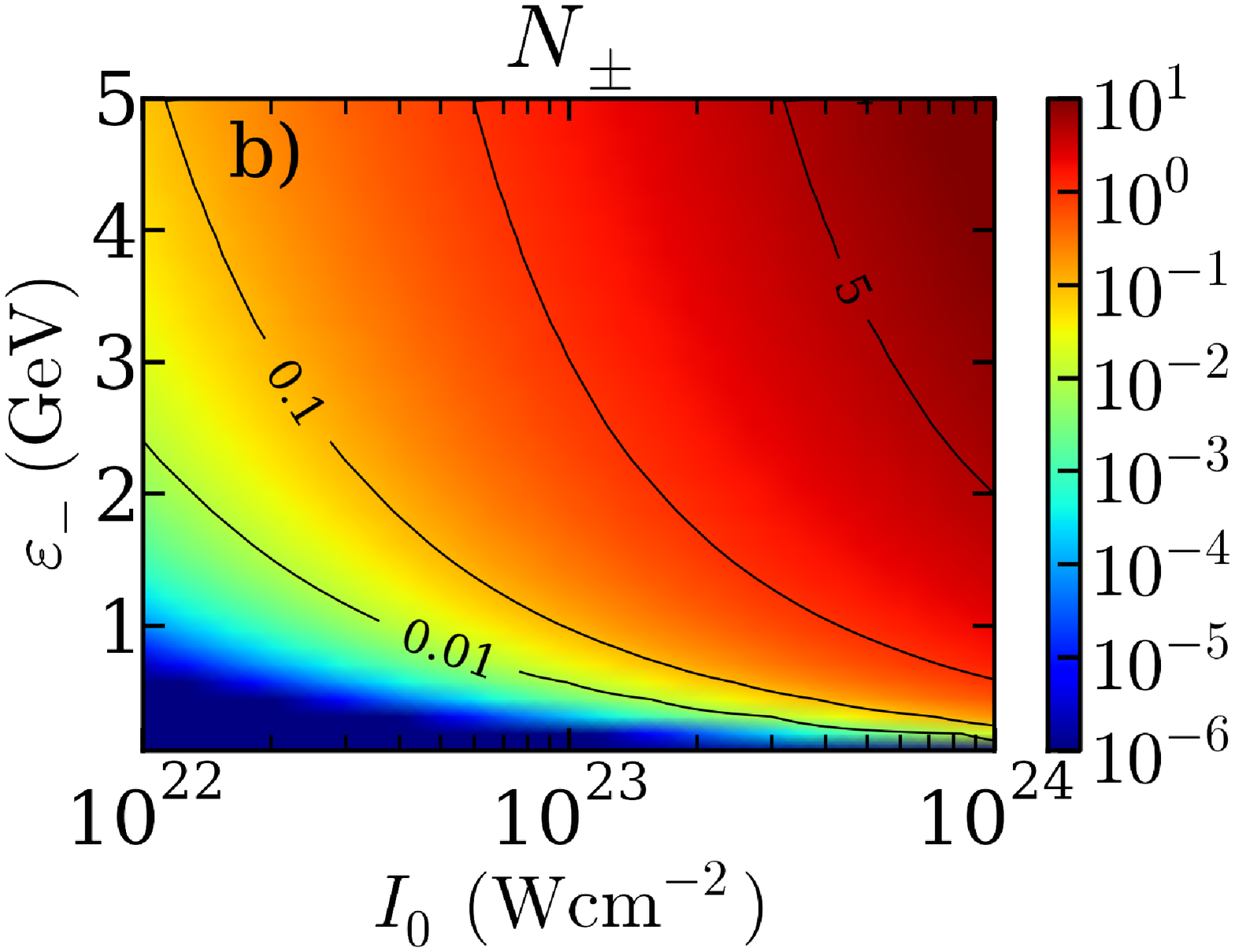}
\caption{Reduced kinetic-QED calculation: number of $\gamma$-ray photons (of energies $\varepsilon_\gamma > 2 m_e c^2$) (a) and positrons (b) produced per laser period
in the collision of a beam electron with a counter-propagating laser wave, as a function of the electron energy and the laser intensity. Countour lines are shown
as black curves.
\label{fig:estimations}} 
\end{center}
\end{figure}

\textit{Simple estimates of pair production} - We start by estimating the efficiency of pair production during the head-on collision of a relativistic electron beam with an intense
laser by means of a reduced kinetic quantum electrodynamic (QED) model \cite{SOKOLOV_PRL_105_195005_2010,ELKINA2011,LOBET_ARXIV_1311.1107_2013}.
This model describes the time evolution of the electron, positron and photon energy distributions under the action of a counter-propagating laser plane wave, taking into
account the nonlinear Compton scattering and Breit-Wheeler pair creation. Unidirectional propagation of the particles at the speed of light is assumed, while advection
and collective effects are neglected. Figures \ref{fig:estimations}(a,b) display, as a function of the laser intensity ($I_0$) and the electron beam energy ($\varepsilon_-$),
the predicted number of photons (of energies $\varepsilon_\gamma > 2m_e c^2$, where $m_e$ is the electron mass) and positrons created per beam electron and
per laser period. In the case of a LWFA driven by the 1-PW laser pulse of the CILEX-Apollon system ($0.8\,\mu\mathrm{m}$ wavelength, $15\,\mathrm{J}$ energy,
$30\,\mathrm{fs}$ FWHM duration, $23\,\mu\mathrm{m}$ FWHM spot size), we expect from Lu's model \cite{Lu_PRSTAB_10_061301_2007} an initial beam energy
$\sim 2\,\mathrm{GeV}$ for a total charge $\sim 1\,\mathrm{nC}$. The beam is then made to collide head-on with the CILEX-Apollon 5-PW, $15\,\mathrm{fs}$ laser
pulse, focused to a maximum intensity of $10^{23}\,\mathrm{Wcm}^{-2}$. Our calculation indicates that each beam electron will produce approximately 20 $\gamma$-ray
photons and 0.5 $e^-e^+$ pair. The interaction will thus yield a positron beam of total charge close to that of the incident electron beam, and of several hundred MeV average
energy (not shown), comparable to the average photon energy.


\textit{Integrated kinetic simulations} - For a more accurate description of the pair production, we make use of two complementary particle-in-cell (PIC) codes.
The laser wakefield acceleration is simulated by means of the quasi-axisymmetric code \textsc{calder-circ} \cite{LIFSCHITZ2009}, based on a reduced cylindrical
discretization of the Maxwell equations, which allows for handling of the spatio-temporal scales of the problem at a reduced computational cost. The resulting
electron beam is then transferred to the three-dimensional (3D) Cartesian code \textsc{calder} \cite{LEFEBVRE_NF_43_629_2003}, enriched with Monte Carlo models
of the photon and pair production \cite{LOBET_ARXIV_1311.1107_2013}. 

\begin{figure}[t]
\begin{center}
\includegraphics[width=0.23\textwidth]{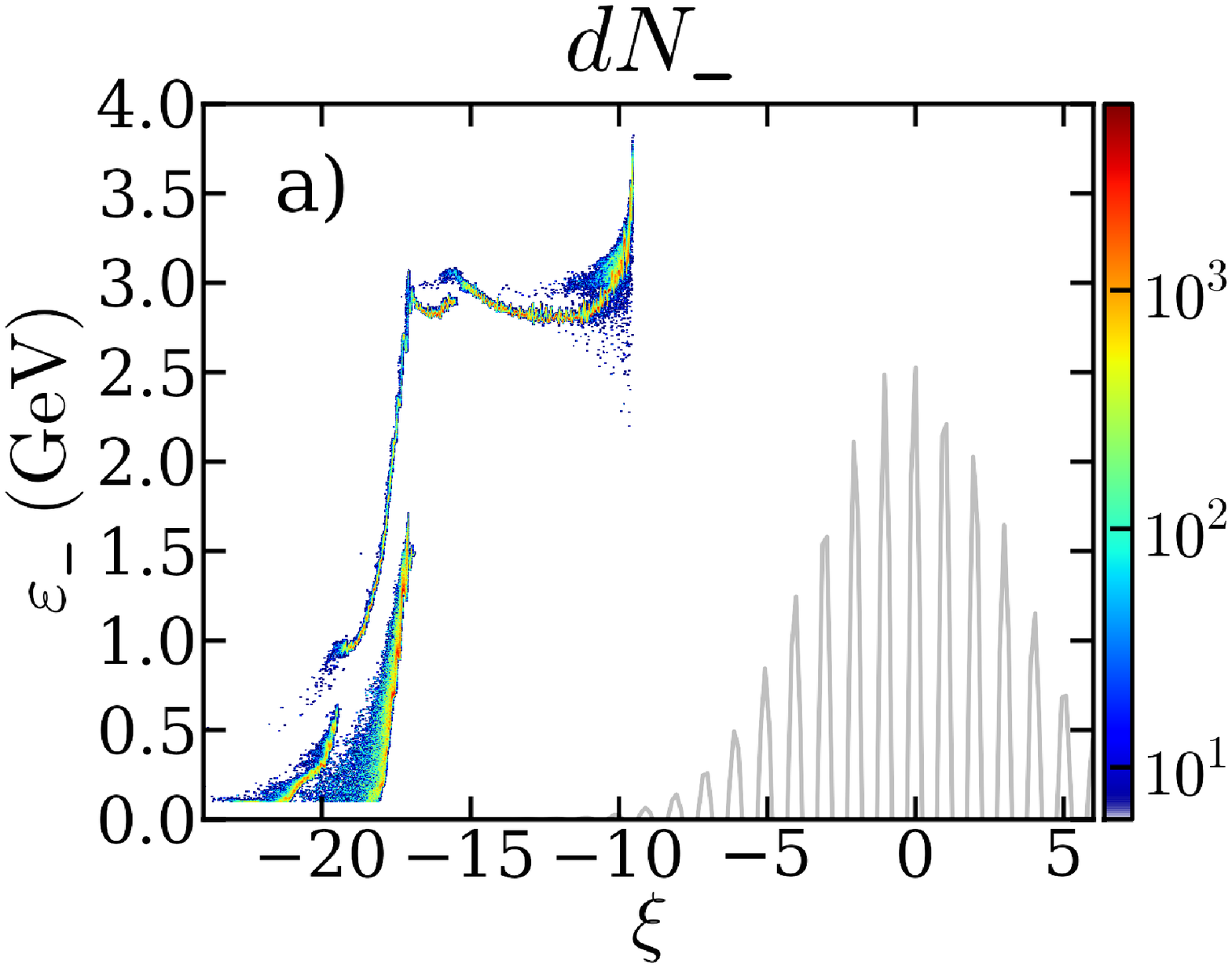}
\includegraphics[width=0.23\textwidth]{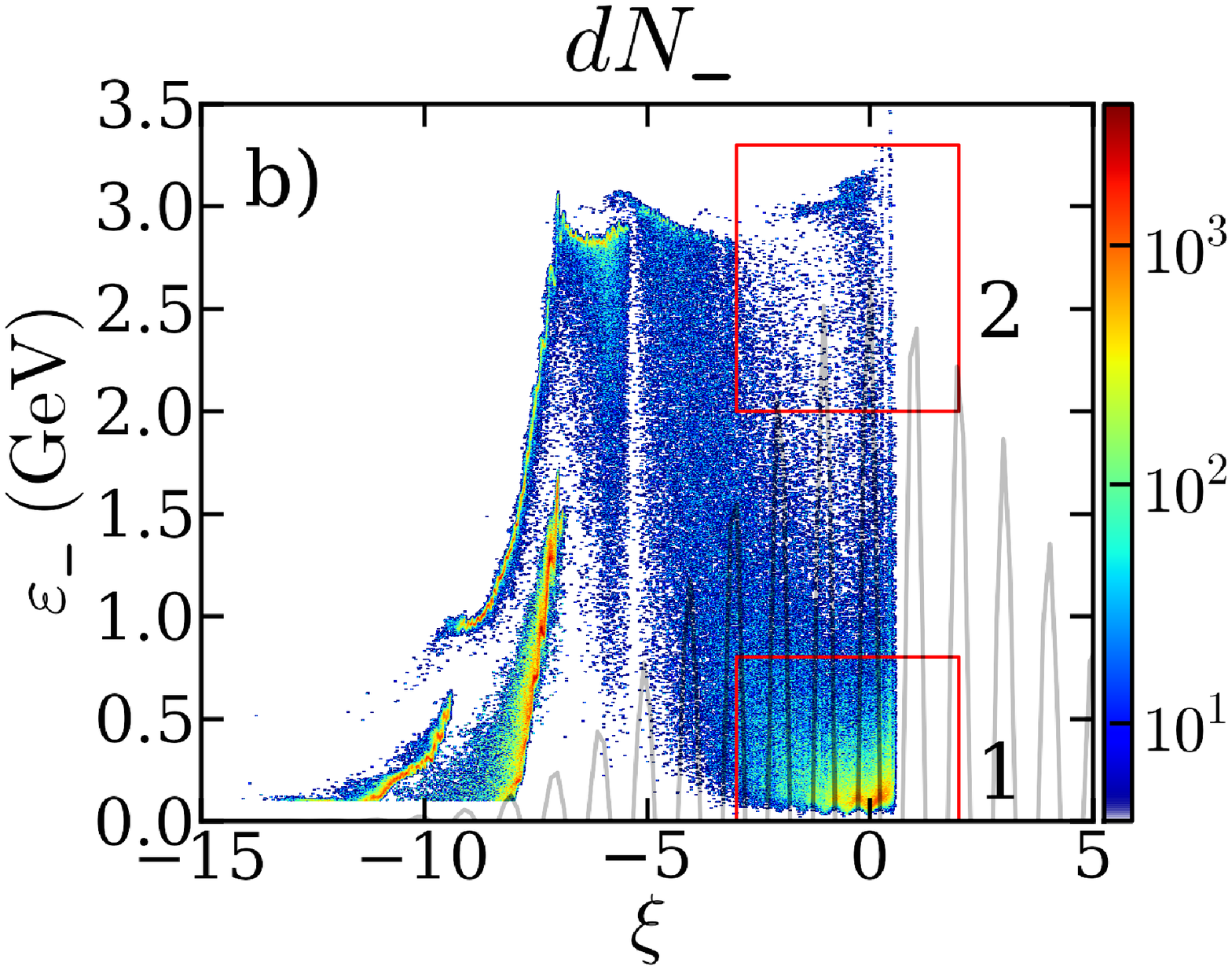}
\includegraphics[width=0.23\textwidth]{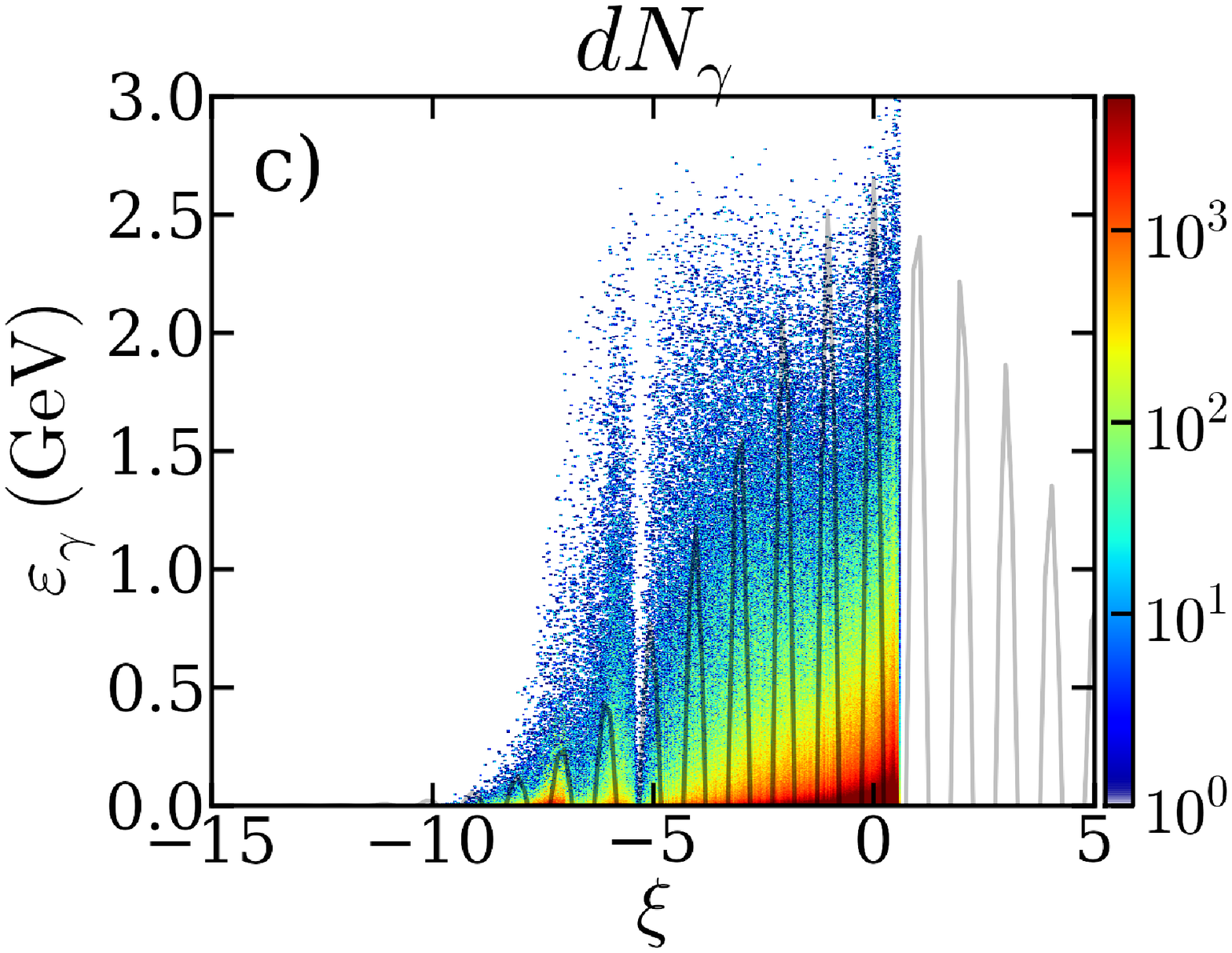}
\includegraphics[width=0.23\textwidth]{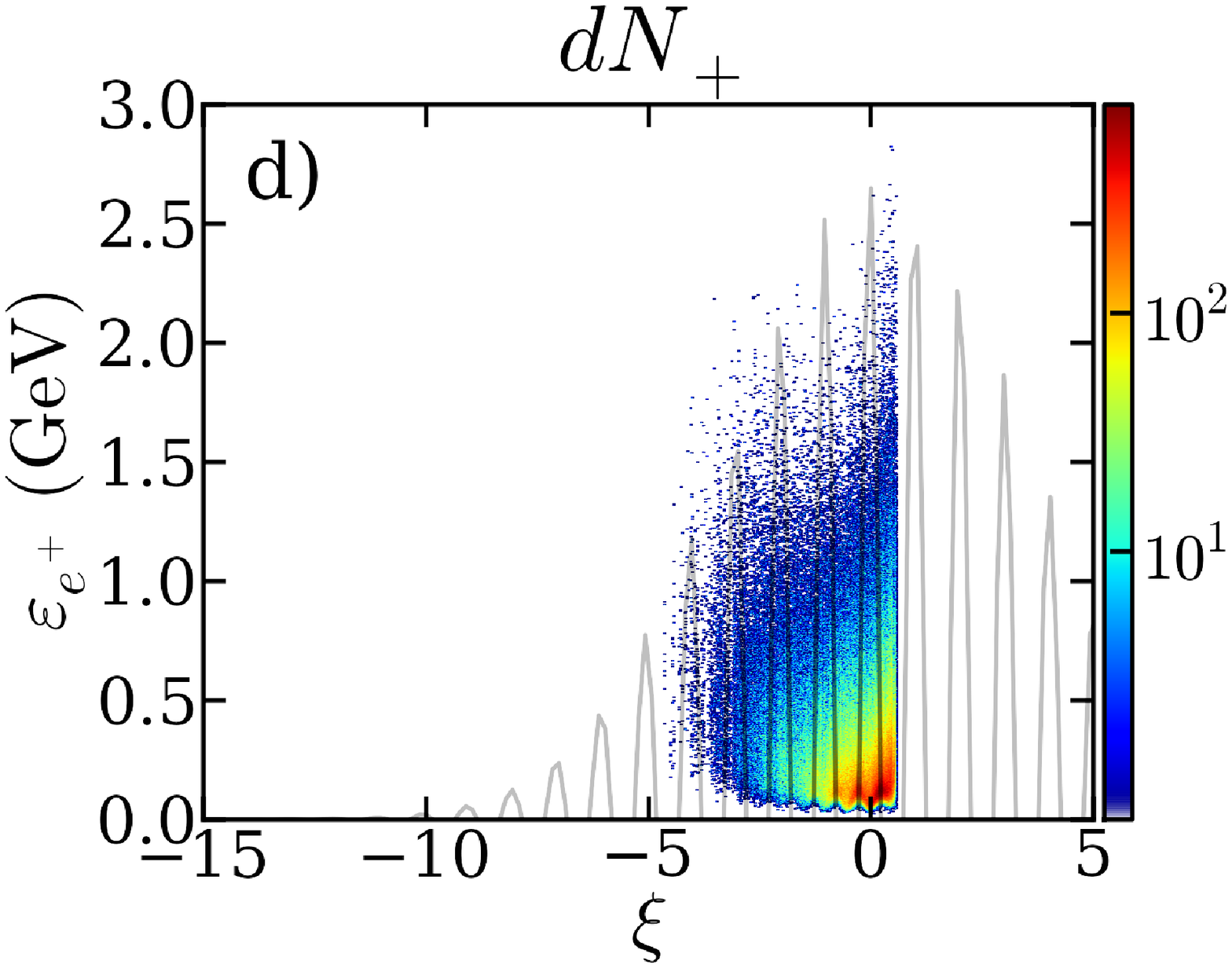}
\includegraphics[width=0.4\textwidth]{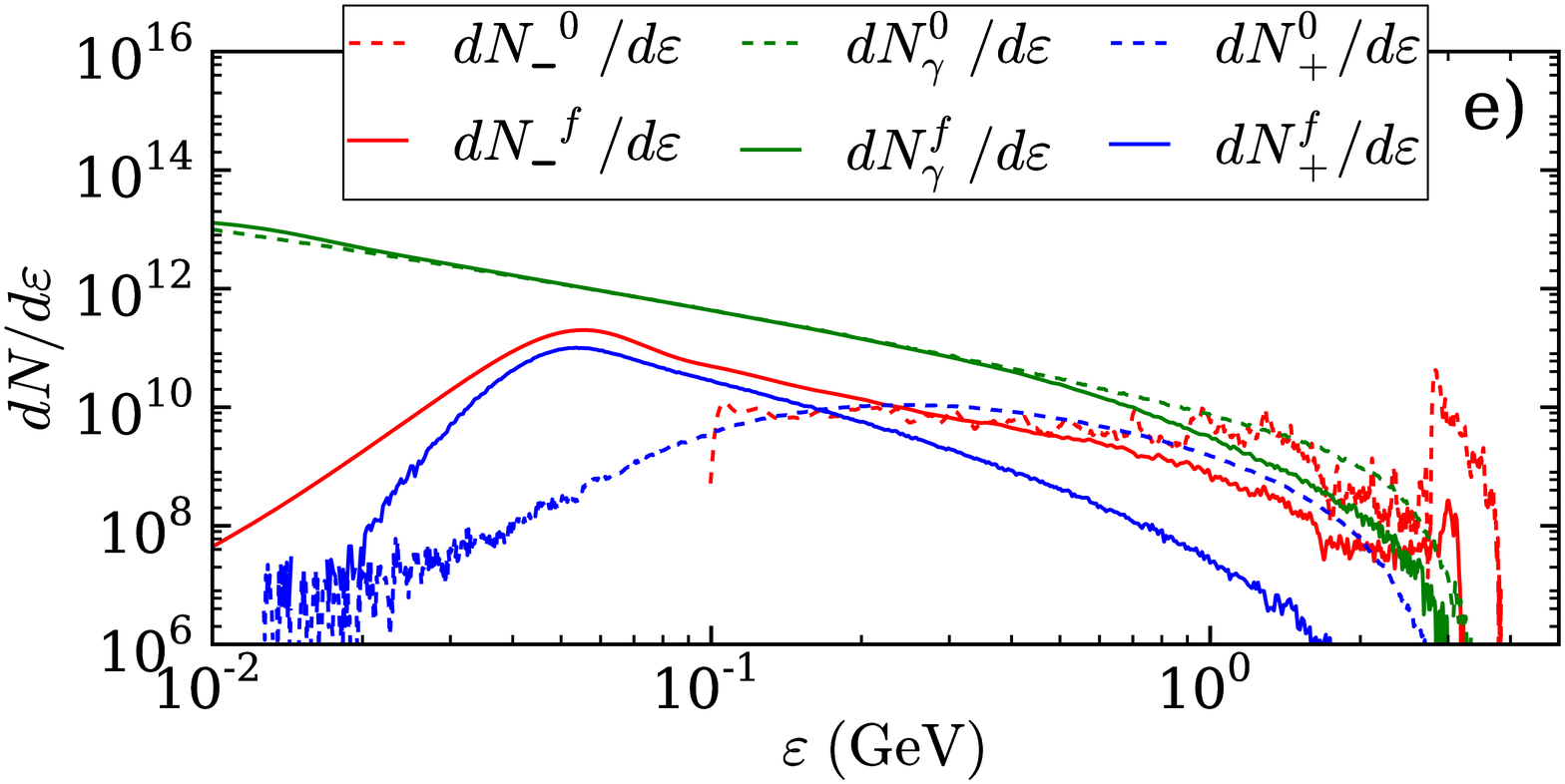}
\caption{(a,b) Electron energy spectra (transversely averaged) before (a) and during (b) the laser-beam collision as a function of the distance
$\xi=\left[x-x_p(t)\right]/\lambda _0$ from the laser peak. In panel (b), frame 1 locates the on-axis beam head after strong radiative deceleration in
the rising part of the laser, while frame 2 corresponds to the beam electrons traveling outside of the laser focus. (c) Photon and (d) positron
energy spectra (transversely averaged) at the time of panel (b). In panels (a-d), the gray line plots the laser $\vert E_y \vert$ profile.
(e) Integrated energy spectra of the electrons (blue), photons (green) and positrons (green) before ($dN^0/d\varepsilon$, dashed lines) and after
($dN^f/d\varepsilon$, solid lines) after the laser-beam collision}
\label{fig:energies}
\end{center}
\end{figure}

\textit{Electron acceleration with the Apollon 1-PW laser} - In a first stage, we simulate the wakefield acceleration induced by the Apollon 1-PW laser
(the parameters of which are given above). We first consider a $1.5\,\mathrm{cm}$-long, flat plasma profile with a density of $0.0016n_c$ ($n_c$ is the
critical density at $1\,\mu\mathrm{m}$). In good agreement with Lu's scaling laws \cite{Lu_PRSTAB_10_061301_2007}, we obtain a 2-GeV electron bunch accompanied
by a broad low-energy tail, with a total beam charge of $5\,\mathrm{nC}$. In order to boost the beam energy, which is a key parameter for pair creation,
we resort to a two-step plasma profile, comprising a $6.3\,\mathrm{mm}$-long plateau at $0.0016n_c$ followed by a $5.6\,\mathrm{mm}$-long plateau at $0.0032n_c$.
This density jump aims at narrowing the plasma bubble when the trapped electrons approach their dephasing length, hence relocating them in the highest
accelerating region of the wakefield, at the back of the bubble. This selects the beam head, increasing its energy up to $3.8\,\mathrm{GeV}$, while reducing its
charge to $\sim 2\,\mathrm{nC}$. The increased monochromaticity of the whole electron beam (above $100\,\mathrm{MeV}$) is clearly seen in the phase space of
\mbox{Fig. \ref{fig:energies}(a)} and the energy spectrum of \mbox{Fig. \ref{fig:energies}(e)}. It is found that 63\% of the total beam charge is contained
in the beam head (between 2.5 and $3.8\,\mathrm{GeV}$). The rest is mostly carried by a secondary electron bunch (with energies under $1.7\,\mathrm{GeV}$)
resulting from self-injection in the shortened bubble. The beam has an average divergence of $3\,\mathrm{mrad}$, transverse FWHM sizes of $4\,\mu\mathrm{m}$
and $2\,\mu\mathrm{m}$ FWHM respectively along and normal to the laser polarization direction ($y$-axis), and a longitudinal length of $12\,\mu\mathrm{m}$.



\begin{figure}[t]
\begin{center}
\includegraphics[width=0.4\textwidth]{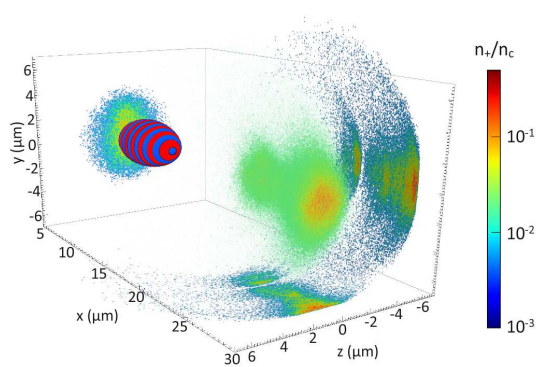}
\caption{Positron density isosurfaces at $t=32\,\mathrm{fs}$ after the laser peak. Density slices along the
$x=27.5\,\mu\mathrm{m}$ (corresponding to the denser part of the beam head), $y=0$ and $z=0$ planes are projected onto the $yz$, $xz$ and $xy$
domain boundaries, respectively. The laser is represented through isosurfaces of the normalized electric field ($eE_y/m_ec\omega_0=\pm 100$).
\label{fig:densities}}
\end{center}
\end{figure}

\textit{Pair generation during the collision with the 5-PW laser} - 
The electron beam is then made to collide head-on with the CILEX-Apollon 5-PW laser. The pulse parameters ($15\,\mathrm{fs}$ duration, $2\,\mu\mathrm{m}$
focal spot) are chosen so as to reach a maximum intensity of $10^{23}\,\mathrm{Wcm}^{-2}$ at the beam head. Consistently with our simple estimates,
the laser-beam interaction is strong enough to cause copious pair production, as depicted in \mbox{Fig. \ref{fig:densities}}. During the collision,
85 \% of the beam energy is radiated in the form of a broad photon distribution [solid green line in \mbox{Fig. \ref{fig:energies}(e)}] extending
from hard x-rays to $\gamma$-rays, with a maximum energy of $3.5\,\mathrm{GeV}$, close to the maximum electron energy, and an average energy of
$33\,\mathrm{MeV}$. Around this energy, the forward-directed photons present an angular divergence of $\sim 3\,\mathrm{mrad}$ (equal to the beam's),
yielding an approximate brilliance of $\sim 5\times 10^{23}\,$photons$\,$s$^{-1}\,$mm$^{-2}\,$mrad$^{-2}\,$0.1\%$\,$BW, three orders of magnitude larger
than the current experimental record \cite{SARRI_PRL_113_224801_2014}.    

Figure \ref{fig:densities} shows that pair production is maximal at the beam head where the tightly focused pulse encounters the most energetic
electrons. The $e^-e^+$ pairs carry $\sim 5$\% of the incident electron beam energy, with a total positron charge $Q_+\sim 0.8\,\mathrm{nC}$ amounting to
$\sim 38\,\%$ of the initial beam charge. As predicted by the reduced QED model, the pulse intensity is too low to achieve global quasi-neutrality (\emph{i.e.},
the production of more than a pair per incident electron). Yet, at the end of the interaction, most of the incident electrons have been expelled by the
laser out of the central pair-filled region, so that the local positron density $n_+ \sim 0.13 n_c$ (averaged over the transverse $\sim 1-2\,\mu\mathrm{m}$
FWHM at the beam head) makes up for a significant fraction ($\sim 40 \%$) of the total leptonic density ($n_-+n_+ \sim 0.32 n_c$). The pair
generation process is detailed in \mbox{Figs. \ref{fig:energies}(a-d)}. Radiative losses become strong (associated with a quantum parameter \cite{DIPIAZZA_RMP_84_1177_2012}
$\chi_- \sim 5\times 10^{-6} a_0\gamma_- \gtrsim 0.1$, where $a_0$ is the normalized wave vector and $\gamma_-$ the electron Lorentz factor) as soon
as the GeV electrons experience a laser intensity $\gtrsim 10^{21}\,\mathrm{Wcm}^{-2}$. Figure \ref{fig:energies}(b) shows that the beam head electrons
have lost more than 90\% of their initial energy by the time they see the laser maximum. Following the interaction, the electron beam spectrum is
strongly broadened towards lower energies, with an average energy of $\sim 190\,\mathrm{MeV}$ [\mbox{Fig. \ref{fig:energies}(e)}, blue curves].
The relatively large value of the maximum energy  ($\sim 3.2\,\mathrm{GeV}$) corresponds to off-axis electrons weakly interacting with the laser field.
This maximum energy is found to drop to $1.5\,\mathrm{GeV}$ for a laser plane wave of the same intensity and duration (not shown).

From comparison of the photon spectra at birth and after the collision [\mbox{Fig. \ref{fig:energies}(e)}, green curves], we find that pair production mainly
arises from photons of energy $\gtrsim 0.5\,\mathrm{GeV}$. These photons have a decay length close to the pulse length, and they are produced in
large numbers only during a few periods before the intensity peak [\mbox{Fig. \ref{fig:energies}(c)}].

The $e^-e^+$ pairs are created with a broad energy distribution (from $\sim 10\,\mathrm{MeV}$ to $3\,\mathrm{GeV}$), associated with an average value
$\sim 0.5\,\mathrm{GeV}$ [Fig. \ref{fig:energies}(e), dashed blue curve]. However, while subsequently moving through the remaining part of the pulse,
they undergo significant radiative losses and the emitted photons can in turn trigger a short-duration pair cascade. At the end of the interaction,
about $4.5\,\%$ of the total positron number originates from cascading, while the positron spectrum is strongly shifted to the low-energy side, with a much
reduced mean energy $\sim 110\,\mathrm{MeV}$ [Fig. \ref{fig:energies}(e), solid blue curve]. 

The final angular spread of the positron beam (\mbox{$\theta^f_{+,xy}\sim 0.33\,\mathrm{rad}$} and $\theta^f_{+,xz}\sim 0.2\,\mathrm{rad}$ in the laser
polarization and perpendicular planes, respectively) exceeds by two orders of magnitude that of the incident electrons ($\sim 0.003\,\mathrm{rad}$), and it
is also significantly larger than the angular spread of the positrons at birth ($\theta^0_{+,xy}\sim 0.014\,\mathrm{rad}$ and $\theta^0_{+,xz}\sim 0.001\,\mathrm{rad}$).
Moreover, it appears to be close to the final divergence of the electron beam ($\theta_{-,xy} \sim \theta_{-,xz} \sim 0.3\,\mathrm{rad}$). As for the electrons,
the increase in the positron angular spread during the laser interaction has two possible origins. First, in the quantum radiation reaction regime considered here,
photon emission in the laser field goes along with strong angular scattering, even in the plane-wave case \cite{WANG_PHP_22_093103_2015}. This deflection
mainly takes place in the $xy$ laser polarization plane: as time increases, the radiating particles progressively lose $x$-momentum while their $y$-momentum
saturates at a value $p_y \propto a_0$ after a few laser cycles \cite{YOFFE_NJP_17_053025_2015,WANG_PHP_22_093103_2015}. This feature starkly contrasts with classical radiation reaction,
through which a particle cannot gain transverse momentum from a plane wave. Second, the particles can be deflected by the transverse ponderomotive force associated
with the small laser spot size. Contrary to the quantum photon emission, which increases the transverse particle momentum only along the polarization $y$-axis,
the ponderomotive force acts along both the $y$- and $z$-axes. The observed close variations of $\theta_{xy}$ and $\theta_{xz}$ during the interaction
for both the positrons and electrons thus demonstrate that ponderomotive and QED effects become comparable under the present conditions (even though QED scattering
still dominates for the positrons). One should note, however,  that the important radiative deceleration in the $x$-direction amplifies the ponderomotive-driven
angular deflection.


\textit{Influence of the laser parameters} - We now examine how the laser parameters affect the positron production. To this goal, we have performed four
laser-electron collision simulations with the same laser duration and energy as above but with varying spot size ($d=2$, 3, 4 and $5\,\mu \mathrm{m}$).
The corresponding peak intensities are $I_0 = 10^{23}$, $4.4 \times 10^{22}$, $2.5 \times 10^{22}$ and $1.6 \times 10^{22}\, \mathrm{Wcm}^{-2}$. To further
assess the influence of the transverse field gradients, this set of simulations is complemented by four simulations employing a laser plane wave at intensities
$I_0 = 10^{23},\ 5 \times 10^{22},\ 2.5 \times 10^{22}$ and $1.25 \times 10^{22}\, \mathrm{Wcm}^{-2}$.

\begin{figure}[t]
\begin{center}
\includegraphics[width=0.23\textwidth]{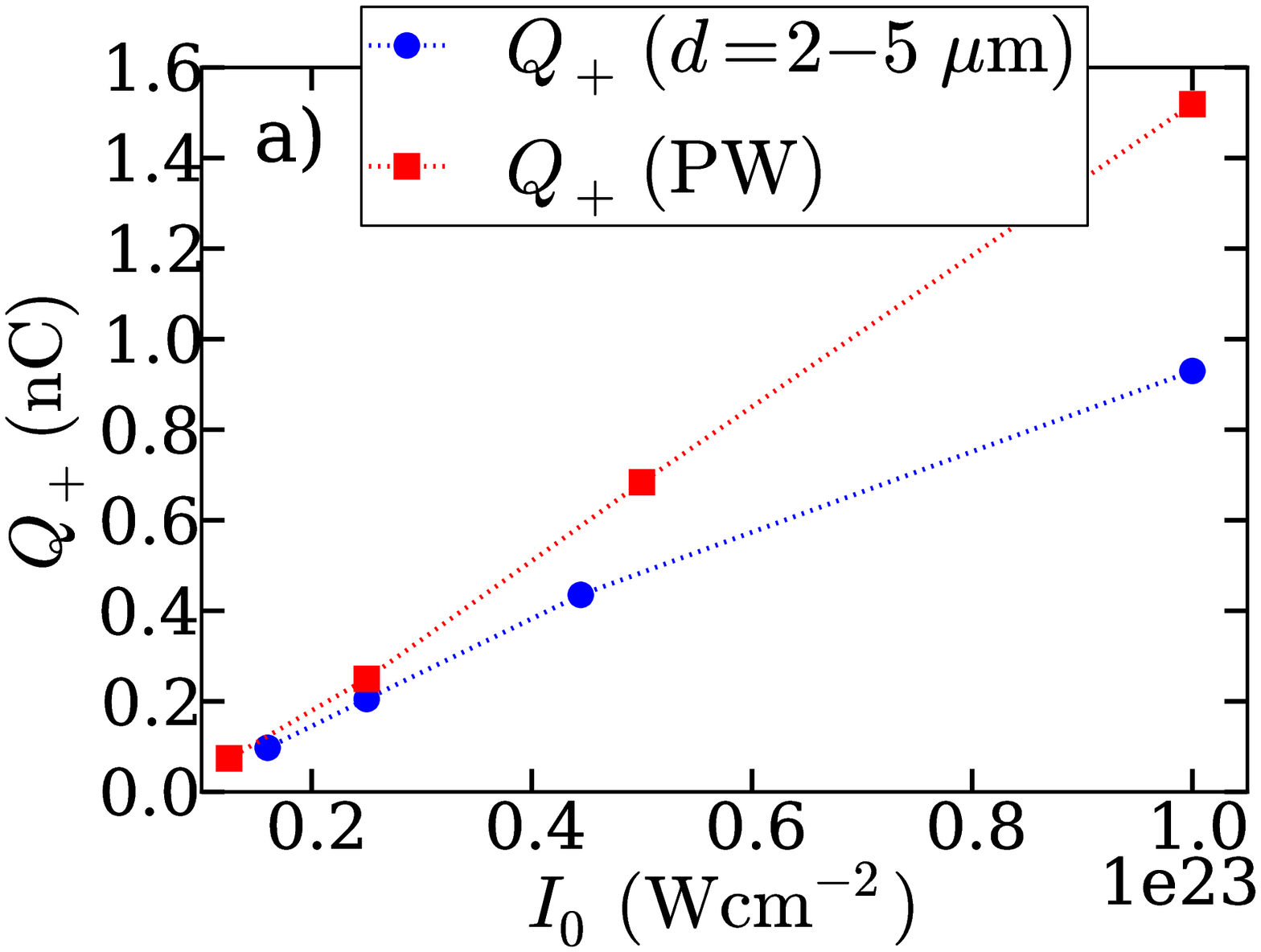}
\includegraphics[width=0.23\textwidth]{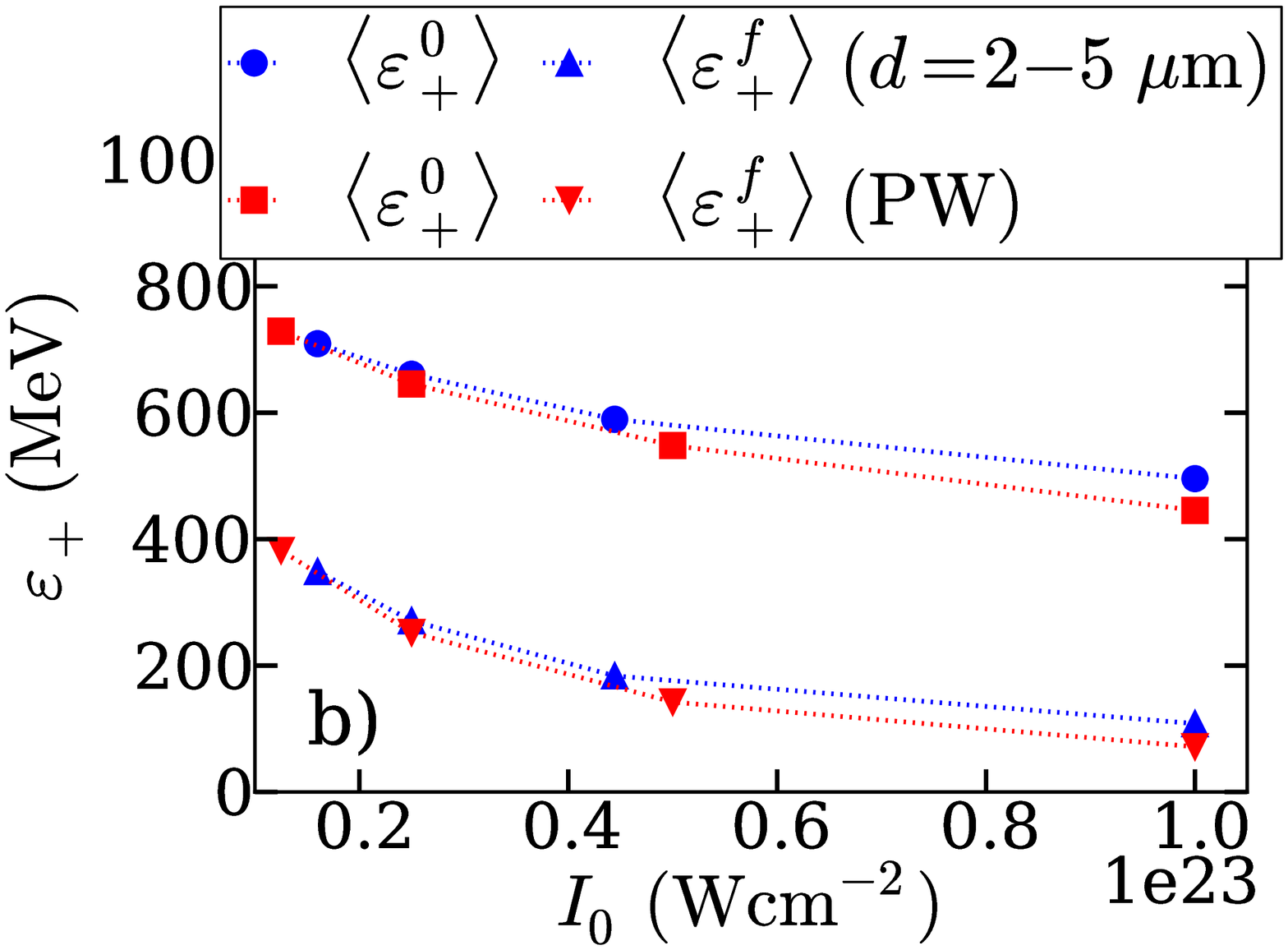}
\includegraphics[width=0.23\textwidth]{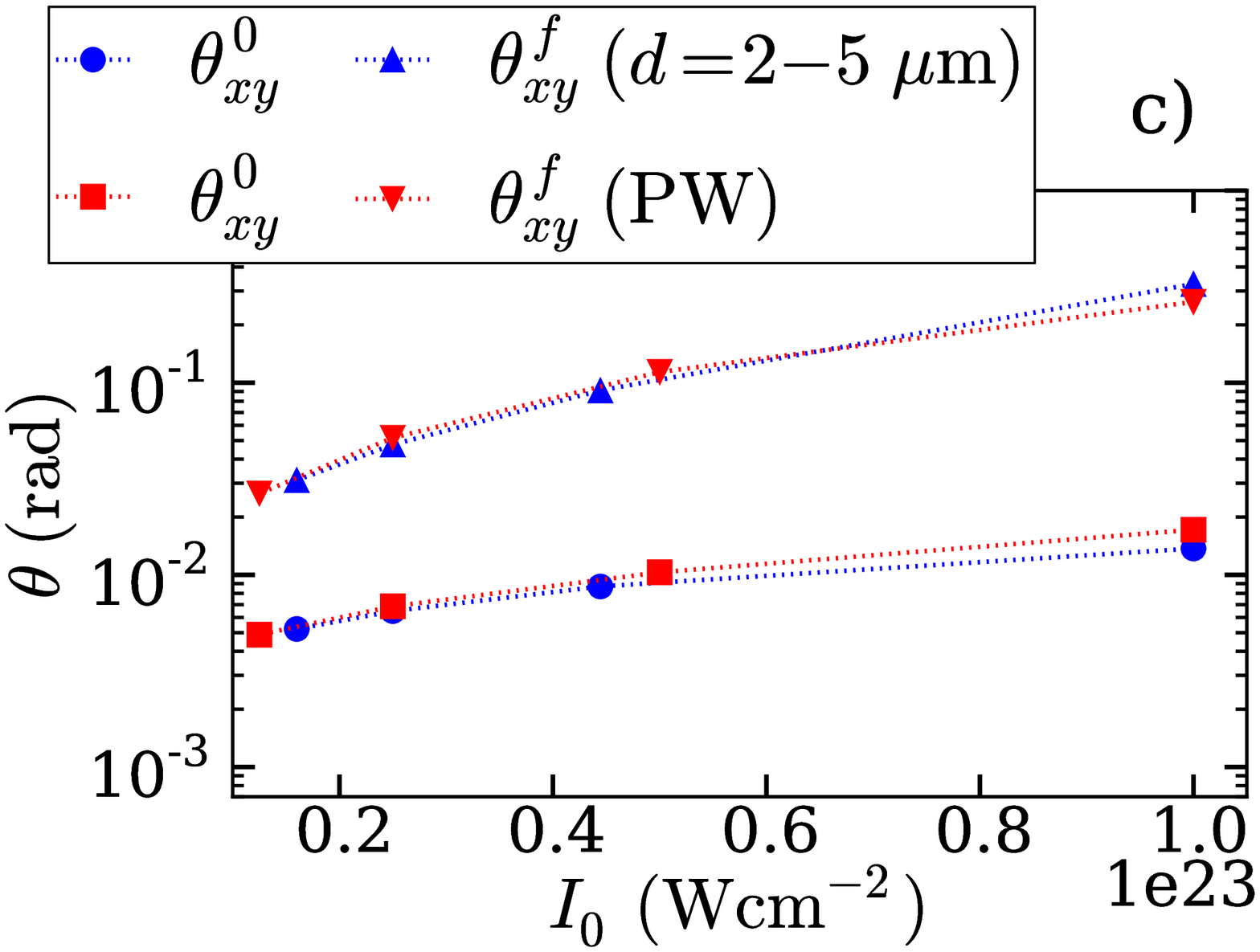}
\includegraphics[width=0.23\textwidth]{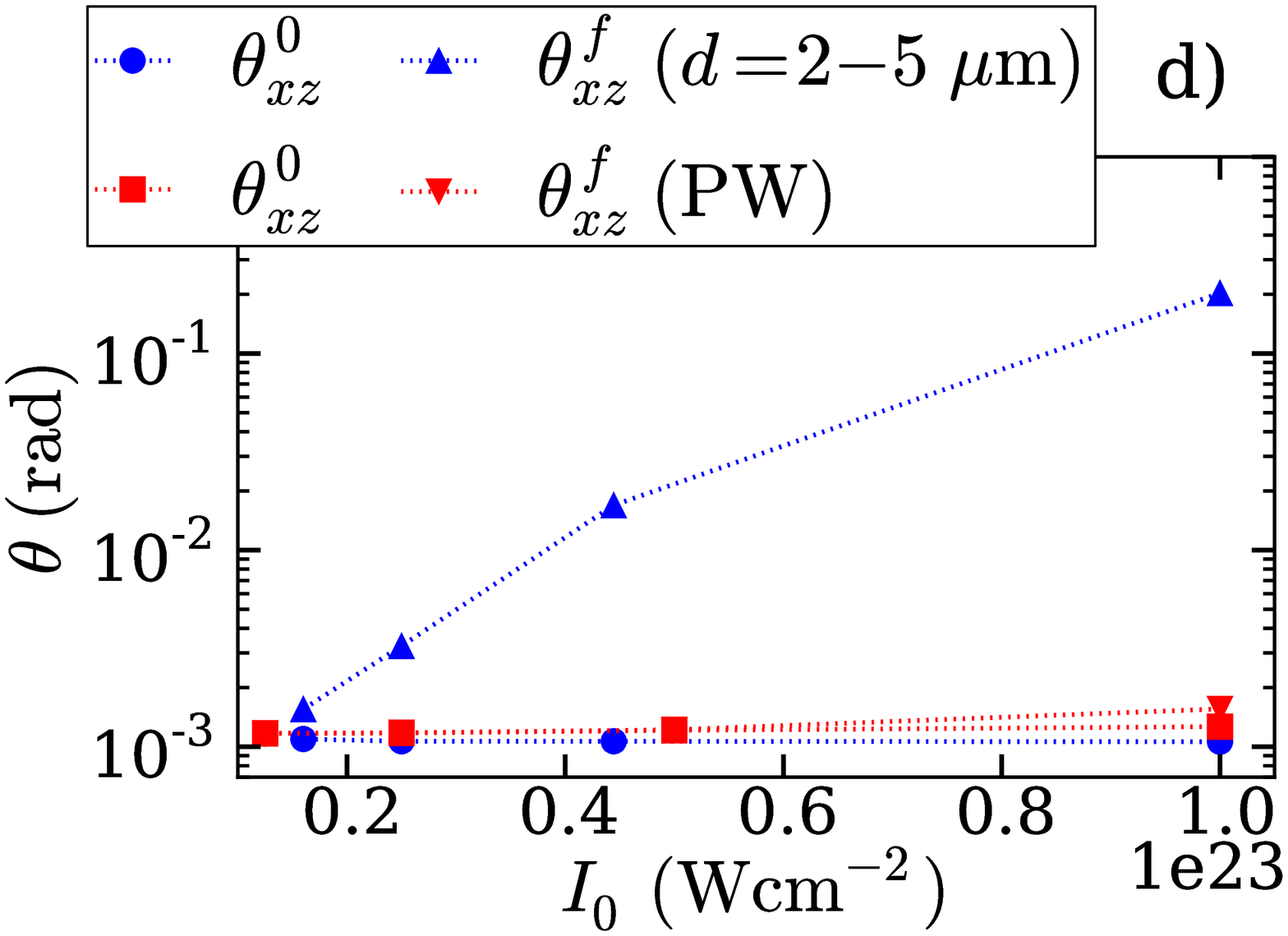}
\caption{Laser intensity dependence of the total positron charge (a), average energy (b) and divergence in the polarization (c) and perpendicular (d) planes
at creation time (square and down triangle markers) and at the end of the interaction (circle and up triangle markers), for a focused wave (blue) or a plane wave (PW, red).
\label{fig:parametric}}
\end{center}
\end{figure}

Figure \ref{fig:parametric}(a) shows that, in the focused case, the total positron charge rises with the laser intensity, albeit at an increasingly slower
rate owing to the dropping number of electrons experiencing a strong field. In the plane-wave case, the positron charge grows linearly as
$Q_+[\mathrm{nC}] \sim 0.17 I_{22}$ (where $I_{22}$ is the intensity in $10^{22}\,\mathrm{Wcm}^{-2}$).  A less expected result (not shown) is that the
number of $\gamma$-ray photons (of energies $> 2m_ec^2$) weakly varies in our range of intensities. This is due to the strong $\gamma$-ray emission
occurring in the foot of high-intensity pulses, which results in important electron deceleration at the time of peak intensity, and therefore in the
production of relatively low-energy photons on average. At lower intensities, radiative losses are diminished so that the $\gamma$-ray emission
takes place over the whole pulse duration at $I_0 = 1.25\times 10^{22}\ \mathrm{Wcm}^{-2}$. Higher intensities, however, augment the probability for
lower-energy photons to decay into pairs and consequently boost the positron yield.

Somewhat surprisingly, Fig. \ref{fig:parametric}(b) reveals that the average positron energy at birth decreases with the laser intensity as
$\langle \varepsilon^0_+ \rangle [\mathrm{MeV}] \sim 490 I_{22}^{-0.8}$ for both the focused and plane waves.
As the intensity rises, low-energy photons can more easily decay into pairs, thus enhancing the number of low-energy positrons and, correspondingly,
lowering the average positron energy. Furthermore, during their subsequent interaction with the laser field, the positrons radiate a larger
fraction of their energy at higher intensities: from $48\,\%$ at $1.25\times 10^{22}\,\mathrm{Wcm}^{-2}$ to $84\,\%$ at $10^{23}\,\mathrm{Wcm}^{-2}$
for a plane wave. This effect further contributes to the decrease in the final mean positron energy with laser intensity.

The intensity dependence of the positron angular spread in the polarization and perpendicular planes is depicted in Figs. \ref{fig:parametric}(c,d).
In the $xy$ polarization plane, the angular spreads in the focused and plane-wave regimes present a similar increase with intensity (from
$\theta_{+,xy}^f\sim 0.025\,\mathrm{rad}$ at $I_0= 1.5\times 10^{22}\,\mathrm{Wcm}^{-2}$ to $\sim 0.3\,\mathrm{rad}$ at $I_0=10^{23}\,\mathrm{Wcm}^{-2}$),
which confirms that QED scattering prevails in the considered intensity range. Both regimes also lead to a similar rise in the divergence during
the interaction (from $\theta_{+,xy}^0$ to $\theta_{+,xy}^f$). In the $xz$ perpendicular plane, the divergence mostly stems from the transverse ponderomotive
force. As a result, for a plane wave, the final angular spread, $\theta_{+,xz}^f$, stays close to the initial value. By contrast, $\theta_{+,xz}^f$
increases with narrowing focal spot and increasing intensity, up to a value close to $\theta_{+,xy}^f$ at $I_0=10^{23}\,\mathrm{Wcm}^{-2}$. 


\textit{Non-collinear geometry} - A non-collinear collision geometry ($\theta \ne 180^\circ$) is generally used in experiments to prevent the reflected
light from damaging the optics. For a $3\,\mu\mathrm{m}$ laser focal spot and a $4.4\times10^{22}\,\mathrm{Wcm}^{-2}$ peak intensity, our simulations
show that the total positron charge remains approximately unchanged ($Q_+\approx 0.4\,\mathrm{nC}$) when the collision angle varies from $\theta =180^\circ$ to $150^\circ$.
In the worst-case perpendicular collision ($\theta = 90^\circ$), the positron yield drops down to $Q_+ = 0.03\,\mathrm{nC}$.


\textit{Conclusion} - Using full-scale 3D PIC simulations, we have demonstrated that soon-to-be-operational, multi-PW, multi-beam lasers will enable all-optical,
high-repetition-rate schemes for efficient Breit-Wheeler pair production, which was as yet only accessible to large-scale accelerators. Besides providing a fully
self-consistent modeling of the problem, our work presents important guidelines for future experiments. Our study thus reveals that the positron yield and mean
divergence (resp. mean energy) increase (resp. decreases) with rising laser intensity at fixed laser energy. In particular, we find that a high-energy
($\sim 400\,\mathrm{MeV}$), low-divergence ($\sim 0.02\,\mathrm{rad}$) positron beam of charge $\sim 0.05\,\mathrm{nC}$ can be achieved using a
moderately-intense ($\sim 10^{22}\,\mathrm{Wcm}^{-2}$) laser pulse focused to a $\sim 5 \,\mu\mathrm{m}$ spot. Once magnetically
segregated from the electrons, this beam could serve as an injector source in conventional or optical accelerators. Higher pulse intensities
($\sim 10^{23}\,\mathrm{Wcm}^{-2}$) are required for generating dense ($Q_+\sim 1\,\mathrm{nC}$, $n_+\sim n_c$), quasi-neutral pair plasmas, at the expense,
however,  of an increased divergence ($\gtrsim 0.1\,\mathrm{rad}$) and a reduced mean energy ($\sim 100\,\mathrm{MeV}$).

\textit{Acknowledgements} - The authors acknowledge support by the French Agence Nationale de la Recherche (LABEX PALM-ANR-10-LABX-39, ANR SILAMPA) and interesting
discussions with S.C. Wilks, F. Amiranoff and P. Audebert. The simulations were performed using HPC resources at TGCC/CCRT (Grant No. 2013-052707). We acknowledge
PRACE for awarding us access to TGCC/Curie (Grant No. 2014112576). We thank the CCRT Team for the helpful support.

\bibliographystyle{unsrt}
\bibliography{biblio}

\end{document}